\documentclass{llncs}
\usepackage{amsmath}
\usepackage{amssymb}
\usepackage{algorithm2e}
\usepackage{graphicx}
\usepackage{psfrag}

\newcommand{\parent}{\mbox{\sf p}}
\newcommand{\status}{\mbox{\sf status}}
\newcommand{\level}{\mbox{\sf level}}
\newcommand{\Nlevel}{\mbox{\sf NewLevel}}

\newcommand{\flevel}{\mbox{$\widehat{level_v}$}}
\newcommand{\fparent}{\mbox{$\widehat{parent_v}$}}

\newcommand{\EProg}{{\bf Propag_{End}}}
\newcommand{\CParent}{{\bf P_{Change}}}
\newcommand{\ULevel}{{\bf Level_{up}}}

\newcommand{\RA}{{\sf R_{InitRoot}}}
\newcommand{\RB}{{\sf R_{SafeChangeP}}}
\newcommand{\RC}{{\sf R_{Level++}}}
\newcommand{\RD}{{\sf R_{EndPropag}}}
\newcommand{\RE}{{\sf R_{LevelCorrect}}}
\newcommand{\RF}{{\sf R_{Dynamic}}}

\newcommand{\removepar}[1]{}
\newcommand{\Algo}{\mbox{\bf \small{Dynamic-LoopFree-BFS}}}

\begin{document}
\title{Universal Loop-Free Super-Stabilization}

\author{
L{\'e}lia Blin\inst{4,2}
\and
Maria Potop-Butucaru\inst{1,2,3}
\and
Stephane Rovedakis\inst{4,5}
\and
S\'{e}bastien Tixeuil\inst{1,2}
}

\institute{
Univ. Pierre \& Marie Curie - Paris 6, France\\
\and
LIP6-CNRS UMR 7606, France\\
\email{maria.gradinariu@lip6.fr, sebastien.tixeuil@lip6.fr}
\and
INRIA REGAL, France
\and
Universit\'e d'Evry Val d'Essonne, France\\
\email{lelia.blin@ibisc.univ-evry.fr, stephane.rovedakis@ibisc.univ-evry.fr}
\and
IBISC-EA 4526, France
}


 \maketitle
\begin{abstract}
We propose an univesal scheme to design loop-free and super-stabilizing protocols for constructing spanning trees optimizing any tree metrics (not only those that are isomorphic to a shortest path tree).

Our scheme combines a novel super-stabilizing loop-free BFS with an existing self-stabilizing spanning tree that optimizes a given metric. The composition result preserves the best properties of both worlds: super-stabilization, loop-freedom, and optimization of the original metric without any stabilization time penalty. As case study we apply our composition mechanism to two well known metric-dependent spanning trees: the maximum-flow tree and the minimum degree spanning tree. 
\end{abstract}

\section{Introduction}

\label{sec:intro}

New distributed emergent networks such as P2P or sensor networks face high churn (nodes and links creation or destruction) and various privacy and security attacks that are not easily encapsulated in the existing distributed models. One of the most versatile techniques to ensure forward recovery of distributed systems is that of \emph{self-stabilization}~\cite{Dij74,Dolev00,T09bc}. A distributed algorithm is self-stabilizing if after faults and attacks hit the system and place it in some arbitrary global state, the system recovers from this catastrophic situation without external (\emph{e.g.} human) intervention in finite time. A recent trend in self-stabilizing research is to  complement the self-stabilizing abilities of a distributed algorithm with some additional \emph{safety} properties that are guaranteed when the permanent and intermittent failures that hit the system satisfy some conditions. In addition to being self-stabilizing, a protocol could thus also tolerate crash faults~\cite{GP93c,AH93c}, nap faults~\cite{DW97j,PT97j}, Byzantine faults~\cite{DW04j,BDH08c,MT06cb,MT07j}, a limited number of topology changes~\cite{DH97j,H00j,KUFM02j} and sustained edge cost changes~\cite{CG02j,JT03}. 

The last two properties are especially relevant when building optimized spanning trees in dynamic networks, since the cost of a particular edge and the network topology are likely to evolve through time. If a spanning tree protocol is \emph{only} self-stabilizing, it may adjust to the new costs or network topology in such a way that a previously constructed spanning tree evolves into a disconnected or a looping structure (of course, in the absence of network modifications, the self-stabilization property guarantees that \emph{eventually} a new spanning tree is constructed). Now, a packet routing  algorithm is \emph{loop free}~\cite{Aceves93,Gafni81} if at any point in time the routing tables are free of loops, despite possible modification of the edge-weights in the graph (\emph{i.e.}, for any two nodes $u$ and $v$, the actual routing tables determines a simple path from $u$ to $v$, at any time). The \emph{loop-free} property~\cite{CG02j,JT03} in self-stabilization guarantees that, a spanning tree being constructed (not necessarily a ``minimal'' spanning tree for some metric), then the self-stabilizing convergence to a ``minimal'' spanning tree maintains a spanning tree at all times (obviously, this spanning tree is not ``minimal'' at all times). The consequence of this safety property in addition to that of self-stabilization is that the spanning tree structure can still be used (\emph{e.g.} for routing) while the protocol is adjusting, and makes it suitable for networks that undergo such very frequent dynamic changes. 
In order to deal with the network churn, \emph{super-stabilization} captures the quality of services a tree stucture can offer during and after a localized topological change. Super-stabilization \cite{Dolev_SuperStab} is an extension of self-stabilization for dynamic settings. The idea is to provide some minimal guarantees (a \emph{passage} predicate) while the system repairs after a topology change. In the case of optimized spanning trees algorithms while converging to a correct configuration (\emph{i.e.} an optimized tree) after some topological change, the system keeps offering the tree service during the stabilization time to all members that have not been affected by this modification. 

\paragraph{Related works}

Relatively few works investigate merging self-stabilization and loop free routing, with the notable exception of~\cite{CG02j,JT03,BlinPRT09}. In~\cite{CG02j}, Cobb and Gouda propose a self-stabilizing algorithm which constructs spanning trees with loop-free property. This algorithm allows to optimize general tree metrics from a considered root, such as bandwidth, delay, distance, etc ... To this end, each node maintains a value which reflects its cost from the root for the optimized metric, for example the maximum amount of bandwith on its path to reach the root. The basic idea is to allows a node to select a neighbor as its parent if this one offers a better cost. To avoid loop creation, when the cost of its parent or the edge-cost to its parent changed a propagation of information is started to propagate the new value. A node can safely change its parent if its propagation of information is ended. Thus, a node can not select one of its descendant as its parent. This algorithm requires a upper bound on the network diameter known to every participant to detect the presence of a cycle and to reset the states of the nodes. Each node maintains its distance from the root and a cycle is detected when the distance of a node is higher than the diameter upper bound.

Johnen and Tixeuil~\cite{JT03} propose another loop-free self-stabilizing algorithm constructing spanning trees, which makes no assumption on the network. This algorithm follows the same approach used in~\cite{CG02j}, that is using propagation of information in the tree. As in~\cite{CG02j}, this second algorithm constructs trees optimizing several metrics from a root, \emph{e.g.}, depth first search tree, breadth first search tree, shortest path tree, etc. Since no upper bound on the network diameter is used, when a cycle is present in the initial network state the protocol continues the initiate propagation of information to grow the value of the nodes in the cycle. The values of these nodes grow until the value of a node reaches a threshold which is the value of a node out of the cycle. Thus, the node reaching this threshold discover a neighbor which offers a better value and can select it to break the cycle. When no cycle is present in the network, the system converges to a correct state.

Also, both protocols use only a reasonable amount of memory ($O(\log n)$ bits per node) and consider networks with static topology and dynamic edge costs. However, the metrics that are considered in~\cite{CG02j,JT03} are derivative of the shortest path (distance graph) metric, that is considered a much easier task in a distributed setting than that of tree metrics not based on distances, \emph{e.g.}, minimum spanning tree, minimum degree spanning tree, maximum leaf spanning tree, etc. Indeed, the associated metric is \emph{locally optimizable}~\cite{GS99c}, allowing essentially locally greedy approaches to perform well. By contrast, some sort of \emph{global optimization} is needed for tree metrics not based on distances, which often drives higher complexity costs and thus less flexibility in dynamic networks.

Recently, \cite{BlinPRT09} proposed a loop-free self-stabilizing algorithm to solve the minimum spanning tree problem for networks, assuming a static topology but dynamic edge costs. None of the previously mentioned works can cope with both dynamic edge changes (loop-freedom) and dynamic local topology changes (super-stabilization). Also, previous works are generic only for local tree metrics, while global tree metrics require \emph{ad hoc} solutions.

\paragraph{Our contributions}

We propose a distributed generic scheme to transform existing self-stabilizing protocols that construct spanning tree optimizing an arbitrary tree metric (local or global), adding loop-free and super-stabilizing properties to the input protocol. Contrary to existing generic protocols~\cite{CG02j,JT03}, our approach provides the loop-free property for \emph{any} tree metric (global or local, rather than only local). Our technique also adds super-stabilization, which the previous works do not guarantee. 
Our scheme consists in composing a distributed self-stabilizing spanning tree algorithm (established and proved to be correct for a given metric) with a novel BFS construction protocol that is both loop-free and super-stabilizing. The output of our scheme is a loop-free super-stabilizing spanning tree optimizing the tree metric of the input protocol.
Moreover, we provide complexity analysis for the BFS construction in both static and dynamic settings. We examplify our scheme with two case study: the maximum flow tree and the minimum degree spanning tree. In both cases, the existing self-stabilizing algorithms can be enhanced via our method with both loop-free and super-stabilizing properties. Interestingly enough, the stabilization time complexity of the original protocols is not worsen by the transformation.

\section{Model and notations} 
\label{sec:model}


We consider an undirected weighted connected network $G=(V,E,w)$ where $V$ is the set of nodes, $E$ is the set of edges and $w: E \rightarrow {\mathbb R^+}$ is a positive cost function. 
Nodes represent processors and edges represent bidirectional communication links. Additionally, we consider that $G=(V,E,w)$ is a dynamic network in which the weight of the communication links and the sets of nodes and edges may change. 
We consider anonymous networks (i.e., processors have no IDs), with one distinguished node, called the \emph{root}\footnote{Observe that the two self-stabilizing MST algorithms mentioned in the Previous Work section assume that the nodes have distinct IDs with no distinguished  nodes. Nevertheless, if the nodes have distinct IDs then it is possible to elect one node as a leader in a self-stabilizing manner. Conversely, if there exists one distinguished node in an anonymous network, then it is possible to assign distinct IDs to the nodes in a self-stabilizing manner~\cite{Dolev00}. Note that it is not possible to compute deterministically  a MST in a fully anonymous network (i.e., without any distinguished node), as proved in \cite{GuptaS03}.}. Throughout the paper, the root is denoted $r$. We denote by $\deg(v)$ the number of $v$'s neighbors in $G$. The $\deg(v)$ edges incident to any node $v$ are labeled from 1 to $\deg(v)$, so that a processor can distinguish the different edges incident to a node. 

The processors asynchronously execute their programs consisting of a set of variables and a finite set of rules. The variables are part of the shared register which is used to communicate with the neighbors. A processor can read and write its own registers and can only read the shared registers of its neighbors. 
Each processor executes a program consisting of a sequence of guarded rules. Each \emph{rule} contains a \emph{guard} (boolean expression over the variables of a node and its neighborhood) and an \emph{action} (update of the node variables only). Any rule whose guard is \emph{true} is said to be \emph{enabled}. A node with one or more enabled rules is said to be \emph{privileged} and may make a \emph{move} executing the action corresponding to the chosen enabled rule.

A {\it local state} of a node is the value of the local variables of the node and the state of its program counter. A {\it configuration} of the system $G=(V,E)$ is the cross product of the local states of all nodes in the system. The transition from a configuration to the next one is produced by the execution of an action of at least one node. A {\it computation} of the system is defined as a \emph{weakly fair, maximal} sequence of configurations, $e=(c_0, c_1, \ldots c_i, \ldots)$, where each configuration $c_{i+1}$ follows from $c_i$ by the execution of a single action of at least one node. During an execution step, one or more processors execute an action and a processor may take at most one action. \emph{Weak fairness} of the sequence means that if any action in $G$ is continuously enabled along the sequence, it is eventually chosen for execution. \emph{Maximality} means that the sequence is either infinite, or it is finite and no action of $G$ is enabled in the final global state.

In the sequel we consider the system can start in any configuration. That is, the local state of a node can be corrupted. Note that we don't make any assumption on the bound of corrupted nodes. In the worst case all the nodes in the system may start in a corrupted configuration. In order to tackle these faults we use self-stabilization techniques.

\begin{definition}[self-stabilization]
Let $\mathcal{L_{A}}$ be a non-empty \emph{legitimacy predicate}\footnote{A legitimacy predicate is defined over the configurations of a system and is an indicator of its correct behavior.} of an algorithm $\mathcal{A}$ with respect to a specification predicate $Spec$ such that every configuration satisfying $\mathcal{L_{A}}$ satisfies $Spec$. Algorithm $\mathcal{A}$ is \emph{self-stabilizing} with respect to $Spec$ iff the following two conditions hold:\\
\textsf{(i)} Every computation of $\mathcal{A}$ starting from a configuration satisfying $\mathcal{L_A}$ preserves $\mathcal{L_A}$ (\emph{closure}).  \\
\textsf{(ii)} Every computation of $\mathcal{A}$ starting from an arbitrary configuration contains a configuration that satisfies $\mathcal{L_A}$ (\emph{convergence}).
\end{definition}

We define bellow a \emph{loop-free} configuration of a system as a configuration which contains paths with no cycle between any couple of nodes in the system.


\begin{definition}[Loop-Free Configuration]
Let $Cycle(u,v)$ be the following predicate defined for two nodes $u,v$ on configuration $C$, with $P(u,v)$ a path from $u$ to $v$ described by $C$:
$$Cycle(u,v) \equiv \exists P(u,v), P(v,u) : P(u,v) \cap P(v,u) = \emptyset.$$
A loop-free configuration is a configuration of the system which satisfies $\forall u,v: Cycle(u,v)=false$.
\end{definition}

%

We use the definition of a loop-free configuration to define a \emph{loop-free stabilizing} system.


\begin{definition}[Loop-Free Stabilization]
A distributed system is called loop-free stabilizing if and only if it is self-stabilizing and there exists a non-empty set of configurations 
such that the following conditions hold: \textsf{(i)} Every execution starting from a loop-free configuration reaches a 
loop-free configuration (closure). \textsf{(ii)} Every execution starting from an arbitrary configuration contains a loop-free configuration (convergence).
\end{definition}

\begin{definition}[Super-stabilization~\cite{Dolev_SuperStab}]
A protocol $P$ is super-stabilizing with respect to a class of topology change event $\Lambda$ iff the following two conditions hold:\\
\textsf{(i)} $P$ is self-stabilizing and \textsf{(ii)} for every computation beginning at a legitimate configuration and containing a single topology change events of type $\Lambda$, a passage predicate holds.
\end{definition}

In the sequel we study the problem of constructing a spanning tree optimizing a desired metric in self-stabilizing manner, 
while guaranteeing the loop-free and super-stabilizing properties.

\section{Super-stabilizing Loop-Free BFS}
In this section, we describe the extension of the self-stabilizing 
loop-free algorithm proposed in \cite{JT03} to dynamic networks. 
Furthermore, we disscuss the super-stabilization of new algorithm.
Interestingly, our algorithm preserves the loop-free property without any degradation 
of the time complexity of the original solution. 

\label{sec:dyn_bfs}
\subsection{Algorithm description}

Algorithm \Algo\/ constructs a BFS tree and guarantees the loop-free property for dynamic networks. 
That is, when topological changes arise in the network (add or deletion of nodes or edges) the algorithm maintains a BFS tree 
without creating a cycle in the spanning tree. To this end, each node has two states: \emph{Neutral}, noted $N$, and 
\emph{Propagate}, noted $P$. A node in state $N$ can safely select as parent its neighbor with the smallest distance 
(in hops) from the root without creating a cycle. A node in state $P$ has an incoherent state according to its parent in 
the spanning tree. In this case, the node must not select a new parent otherwise a cycle can be created. 
So, this node has to inform first its descendants in the tree that an incoherency in the BFS tree was detected. Then, it corrects 
when all its subtrees have recovered a coherent state. Therefore, a node $v$ in state $P$ initiates a propagation of 
information with feedback in its subtree§. When the propagation is finished the nodes in the subtree $v$ 
(including $v$) recovers a correct distance and the state $N$.

We consider a particular node $r$ which acts as the root of the BFS tree in the network. Every node executes the same 
algorithm, except the root which uses only Rule $\RA$ to correct its state. 
In a correct state, the root $r$ of the BFS tree has no parent, a zero level and the state $N$. 
Otherwise, Rule $\RA$ is executed by $r$ to correct its state.

The other five rules are executed by the other nodes of the network. 

Rule $\RB$ is used by a node $v$ with the state $N$ 
if it detects a better parent, i.e., a neighbor node with a lower level than the level of its actual parent. 
In this case, $v$ can execute this rule to update its state in order to select a new parent without creating a cycle in the tree.

If a node $v$ has the best parent in its neighborhood but an incoherent level according to its parent, 
then $v$ executes Rule $\RC$ to change its status to $P$ and to initiate a propagation of information 
with feedback which aims to inform its descendants of its new correct level. A descendant $x$ of node $v$ with state 
$N$ with a parent in state $P$ executes Rule $\RC$ to continue the propagation and to take into account its new level. 

When a leaf node $x$, descendant of $v$ in Status $P$ is reached, $x$ stops the propagation by executing Rule $\RD$ 
to change its state to $N$ and to obtain its correct level. The end of propagation is pull up in the tree using Rule $\RD$. 

Rule $\RE$ corrects at node $v$ the variable used to propagate the new level in the tree (variable $\Nlevel_v$) 
if this variable is lower than the actual level of $v$.

Rule $\RF$ deals with the dynamism of the network. This rule is executed by a node $v$ when it 
detects that its parent is no more in the network and it cannot select with Rule $\RB$ a new parent because of 
its level (otherwise it may create a cycle). The aim of this rule is to increase the level of node $v$ using propagations of 
information as with Rule $\RC$, until $v$'s level allows $v$ to select a neighbor as its new parent without creating a cycle.

\begin{figure}[h]
\includegraphics[scale=0.65]{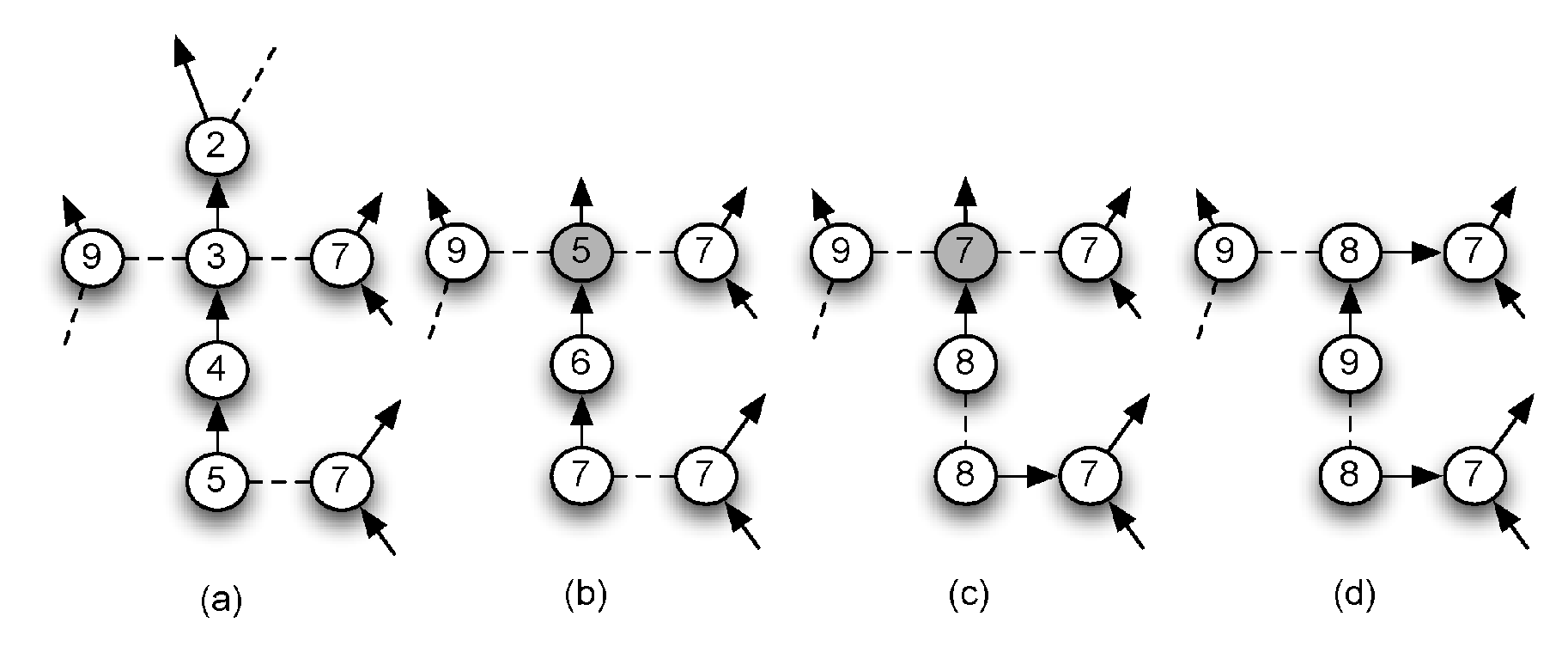}
\caption{Correction of the BFS tree after a node deletion.}
\label{fig:RD}
\end{figure}

Figure~\ref{fig:RD} illustrates the mechanic of Rule $\RF$. In Figure~\ref{fig:RD}(a) is depicted a part of the constructed BFS tree 
before the deletion of the node of level 2. After the deletion of this node, the node $v$ with level 3 executes Rule $\RF$ to increase 
its level (equal to the lowest neighbor level plus one) in order to recover a new parent. 
Figure~\ref{fig:RD}(b) shows the new level of $v$ and the new levels $v$'s descendants when the first propagation is ended. 
However, a level of 5 is not sufficient to allow $v$ to select a new parent, so a second propagation is started by $v$ 
which affects the levels given by Figure~\ref{fig:RD}(c). Note that a descendant of $v$ can leave $v$'s subtree to 
obtain a better level if possible, this can be observed in Figure~\ref{fig:RD}(c). 
Finally, $v$ reaches a state with a level which allows $v$ to execute Rule $\RB$ to select its new parent, 
and $v$'s descendants execute Rule $\RC$ to correct their levels according to $v$'s level. Figure~\ref{fig:RD}(d) shows the new levels computed by the nodes.

\subsubsection{Detailed level description.}

In the following, we describe the variables, the predicates and the rules used by Algorithm \Algo\/.

\paragraph{Variables:}
For any node $v \in V(G)$, we denote by $N(v)$ the set of all neighbors of $v$ in $G$ and by 
$\mathcal{D}_v$ the set of sons of $v$ in the tree. We use the following notations: 
\begin{itemize}
\item $\parent_v$: the parent of node $v$ in the current spanning tree;
\item $\status_v$: the status of node $v$, $P$ when $v$ is in a propagation phase, $N$ otherwise;
\item $\level_v$:  the number of edges from $v$ to the root $r$ in the current spanning tree;
\item $\Nlevel_v$: the new level in the current spanning tree (used to propagate the new level).
\end{itemize}

\begin{figure}[!ht]
\vspace*{-0.9cm}
\fbox{
\begin{minipage}{12cm}

\begin{small}
\begin{description}
\item $\flevel \equiv \left\{ \begin{array}{lll} \min\{\level_u+1: u \in N(v)\} & & \mbox{if } v \neq r\\ 0 & & \mbox{otherwise} \end{array}\right.$
\item $Min_v\equiv \min\{u: u \in N(v) \wedge \level_u=\flevel-1 \wedge \status_u=N\}$
\item $\fparent \equiv \left\{ 
\begin{array}{lll}
 Min_u & & \mbox{if } \exists u \in N(v), \level_u=\flevel-1\wedge \status_u=N \\
\bot & & \mbox{otherwise}
 \end{array}\right.$
\item $\mathcal{D}_v \equiv \{u: u \in N(v) \wedge \parent_u=v \wedge \level_u>\level_v\}$
\item $ubl_v \equiv \left\{ \begin{array}{lll} \min\{\level_u-1: u \in \mathcal{D}_v\} & & \mbox{if } \mathcal{D}_v \neq \emptyset\\ \infty & & \mbox{otherwise} \end{array} \right.$
\item 
\item $\EProg(v) \equiv (\forall u \in \mathcal{D}_v, \status_u=N)$
\item $\CParent(v) \equiv (\flevel<\level_v \vee (\level_v=\flevel \wedge \parent_v \neq \fparent)) \wedge \fparent \neq \bot$
\item $\ULevel(v) \equiv \level_v \neq \level_{\parent_v}+1 \vee (\status_{\parent_v}=P \wedge \level_v \neq \Nlevel_{\parent_v}+1)$

\end{description}
\end{small}
\end{minipage}
}
\caption{Predicates used by the algorithm.}
\label{fig:predicates}
\vspace*{-0.5cm}
\end{figure}

The root of the tree executes only the first rule, named $\RA$, while the other nodes execute the five last rules.

\begin{description}
\item[$\RA:$ \textbf{(Root Rule)}]~\\
\textbf{if} $v=r \wedge (\parent_v \neq \bot \vee \level_v \neq 0 \vee \Nlevel_v \neq 0 \vee \status_v \neq N)$\\ \textbf{then} $\parent_v:=\bot; \level_v:=0; \Nlevel_v:=0; \status_v:=N;$
\end{description}

\begin{description}
\item[$\RB:$ \textbf{(Safe parent change Rule)}]~\\
\textbf{if} $v \neq r \wedge \status_v=N \wedge \CParent(v)$\\ \textbf{then} $\level_v:=\flevel; \Nlevel_v:=\level_v; \parent_v:=\fparent;$
\end{description}

\begin{description}
\item[$\RC:$ \textbf{(Increment level Rule)}]~\\
\textbf{if} $v \neq r \wedge \status_v=N \wedge \parent_v \in N(v) \wedge \neg \CParent(v) \wedge \ULevel(v)$\\ \textbf{then} $\status_v:=P; \Nlevel_v:=\Nlevel_{\parent_v}+1;$
\end{description}

\begin{description}
\item[$\RD:$ \textbf{(End of propagation Rule)}]~\\
\textbf{if} $v \neq r \wedge \status_v=P \wedge \EProg(v) \wedge ubl_v \geq \Nlevel_v$\\ \textbf{then}
$\status_v:=N; \level_v:=\Nlevel_v;$
\end{description}

\begin{description}
\item[$\RE:$ \textbf{(Level correction Rule)}]~\\
\textbf{if} $v \neq r \wedge \Nlevel_v < \level_v$ \textbf{then} $\Nlevel_v:=\level_v;$
\end{description}

\begin{description}
\item[$\RF:$ (Increment level Rule for dynamic networks)]~\\
\textbf{if} $v \neq r \wedge \status_v=N \wedge \parent_v \not \in N(v) \wedge \neg \CParent(v)$\\ \textbf{then} $\status_v:=P; \Nlevel_v:=\flevel;$
\end{description}

\subsection{Correctness proof}
The algorithm proposed in the precedent subsection extends the algorithm of \cite{JT03} to dynamic network topologies. 
When the system is static the correctness of the algorithm directly follows from the results proven in \cite{JT03-RR}. 
In the following, we focus only the case of dynamic topologies, i.e., when nodes/edges of the tree fails or 
nodes/edges are added in the network. Note that in the following, we only study the case of an edge failure. A 
node failure produces the same consequences, i.e., the spanning tree is splitted and some nodes have no parent. 
Moreover, we do not consider edges out of the tree because this does not lead the system in an 
illegitimate configuration. After each fail of node or edge in the tree, we assume the uderlying network is always connected.

In \cite{JT03-RR}, a legitimate configuration for the algorithm is defined by the following predicate 
satisfied by every node $v \in V$: $\mbox{Pr}_v^{\mathcal{LP}} \equiv [(v=r) \wedge (\level_v=0) \wedge (\status_v=N)] \vee [(v \neq r) \wedge (\level_v=\dot{\level_v}) \wedge (\status_v=N) \wedge (\level_v=\level_{\parent_v}+1)]$, with $\forall v \neq r, \dot{\level_v}=\min\{\dot{\level}_u+1: u \in N(v)\}$ defines the optimal level of node $v$.

Note that after a fail of an edge of the tree $T$, Predicate $\mbox{Pr}_v^{\mathcal{LP}}$ is not satisfied anymore. 
The tree $T$ splits in a forest $F$ which 
contains the subtrees of $T$. Let $Orph$ be the set of nodes $v$ such that $\parent_v \not \in N(v)$, note that $r \not \in Orph$. 
The following predicate is satisfied by every node $v \in V, v \not \in Orph$

%
%

$$\mbox{Pr}_v^{\mathcal{LC}} \equiv \left\{ \begin{array}{ll} \level_{\parent_v}+1 \leq \level_v \wedge \Nlevel_v \geq \level_v & \mbox{if } v \neq r \\ \level_r=0 \wedge \status_r=N & \mbox{otherwise} \end{array} \right.$$

We show below that each node with no parent in $F$ starts a propagation of information in its subtree.

\begin{lemma}
\label{lem:dyn_propag}
Let a node $v \in V, v \in Orph$. If $\status_v=N$ and $\CParent(v)=false$ then status $v$ eventually moves to $P$.
\end{lemma}

\begin{proof}
Let $v \in V, v \in Orph$ be a node such that $\status_v=N$ and $\CParent(v)=false$. $v$ can only execute Rules $\RE$ or $\RF$, because $v$ can not execute Rules $\RB, \RC$ and $\RD$ since $\status_v=N, \CParent(v)=false$ and $\parent_v \not \in N(v)$. To change its status from $N$ to $P$, a node $v \in Orph$ must execute Rule $\RF$. Suppose that $v$ does not execute Rule $\RF$. So $v$ can only execute Rule $\RE$. However, after execution of Rule $\RE$ we have $\Nlevel_v:=\level_v$ and the guard of Rule $\RE$ is no more satisfied. Thus, only the guard of Rule $\RF$ is satisfied and $v$ remains enabled until it performs Rule $\RF$. Therefore, the scheduler eventually selects $v$ to perform Rule $\RF$.
\hspace*{\fill}$\Box$
\end{proof}

According to Lemma 9 in \cite{JT03-RR}, a node $v$ such that $\status_v=P$ eventually performs Rule $\RD$ to change its status to $N$. 
In the following, we show that a node in $Orph$ (i.e., without a parent in its neighborhood) eventually leaves the set $Orph$.

\begin{lemma}
\label{lem:dyn_spanning_tree}
Let $v \in V, v \in Orph$. Eventually, $v$ is not anymore in the set $Orph$ and selects a parent without creating a cycle.
\end{lemma}

\begin{proof}
We show the lemma by induction on the height of the subtree of $v$. Consider the case where a node $v \in Orph$ has a neighbor $u \in N(v)$ such that $\level_u<\level_v$. We assume that for every node $x$ in $F$, $x \not \in Orph$, we have $\level_{p_x}+1 \leq \level_x$. So, $u$ can not be a descendant of $v$. Thus, $v$ performs Rule $\RB$ to choose $u$ as its parent without creating any cycle in $F$. Otherwise, every node $u \in N(v)$ is a child of $v$. According to Lemma 9 in \cite{JT03-RR} and Lemma \ref{lem:dyn_propag} (above), the level of every node in the subtree of $v$ increases. Since we assume the network is always connected, there exists a leaf node $x$ in the subtree of $v$ such that $\level_x>\dot{\level}_x=\level_y$, with $y \in N(x)$. Thus, $x$ can execute Rule $\RB$ to choose $y$ as its parent and $x$ leaves the subtree of $v$. Since the height of the subtree of $v$ is finite, eventually $v$ can choose a neighbor $u$ as its parent because $u$ is no more in the subtree of $v$. Therefore, in a finite time a node $v \in Orph$ leaves the set $Orph$ by selecting a parent in its neighborhood without creating a cycle.
\hspace*{\fill}$\Box$
\end{proof}

According to Lemma \ref{lem:dyn_spanning_tree}, each node has a parent and no cycle is created. Thus, the system reaches a configuration where a spanning tree is constructed. So the analysis given in \cite{JT03-RR} can be used to show that the system reaches a configuration in which for each node $v \in V$ we have $\level_v=\dot{\level}_v$. Since the initial configuration contains a spanning tree, the algorithm stabilizes to a breadth first search tree and during the stabilization of the algorithm the loop-free property is maintained, as showed in \cite{JT03-RR}.

Above we consider only the fail of nodes/edges of the tree, now we discuss the add of nodes and edges in the network. In a legitimate configuration, after the add of an edge every node $v \in V$ always satisfies $\level_v \geq \dot{\level}_v$. According to Lemma 12 and Corollary 1 in \cite{JT03-RR}, in a finite time eventually for every node $v \in V$ we have $\level_v=\dot{\level}_v$. In a legitimate configuration, after the add of a node $v$ Rule $\RB$ is executed by $v$ to select a neighbor $u \in N(v)$ as its parent, there exists such a node $u$ because we assume that the network is always connected. Therefore, the system is in an arbitrary configuration where a spanning tree is constructed. Therefore, the analysis given in \cite{JT03-RR} can be used to show that in a finite time for every node $v \in V$ we have $\level_v=\dot{\level}_v$. Moreover, in the case of node/edge adds the initial configuration contains a spanning tree, thus the loop-free property is maintained by the algorithm.\\

In the following, we prove that the presented algorithm has a superstabilizing property for a particular class of topology change events. We show that a passage predicate is satisfied during the restabilizing execution of our algorithm. We define the considered topology change events, noted $\varepsilon$:
\vspace*{-0.2cm}
\begin{itemize}
\item an add (resp. a removal) of an edge $(u,v)$ in the network noted ${\tt recov}_{uv}$ (resp. ${\tt crash}_{uv}$);
\item an add (resp. a removal) of a neighbor node $u$ of $v$ in the network noted ${\tt recov}_u$ (resp. ${\tt crash}_u$).
\end{itemize}

In the sequel, we suppose that after every topology change event the network remains connected. We provide below definitions of the topology change events class $\Lambda$ and passage predicate.

\begin{definition}[Class $\Lambda$ of topology change events]
 ${\tt crash}_{uv}$ and ${\tt crash}_v$ compose the class
 $\Lambda$ of topology change events.
\end{definition}

\begin{definition}[Passage predicate]
\label{def:passage_predicate}
The parent of a node $v$ can be modified if $v$ is in the subtree connected by
the removed edge or node, and the parent is not changed
for any other node in the tree.
\end{definition}

\begin{lemma}
\label{lem:super_stab}
The proposed protocol is superstabilizing for the class $\Lambda$ of
topology change events, and the passage predicate (Definition
\ref{def:passage_predicate}) continues to be satisfied while a
legitimate configuration is reached.
\end{lemma}

\begin{proof}
Consider a legitimate configuration $\Delta$. Suppose $\varepsilon$ is a removal of edge $(u,v)$ from the network. If $(u,v)$ is not a tree edge then the levels of $u$ and $v$ are not modified and neither $u$ nor $v$ changes its parent, thus no parent variable is modified. Otherwise, let $\parent_v=u$, $u$'s level and $u$'s parent are not modified, it is true for any other node $x$ not contained in the subtree of $v$ since the distance between $x$ and the root $r$ in the graph is not modified (i.e., Predicate $\CParent(x)$ is not satisfied). However, $u$ is no more a neighbor of $v$ so according to Lemma~\ref{lem:dyn_propag} $v$ executes Rule $\RF$ and starts a propagation phase. Moreover, according to Lemma~\ref{lem:dyn_spanning_tree} $v$ selects a new parent without creating a cycle. Therefore, only a node in the subtree connected by the edge $(u,v)$ may change its parent.

Suppose $\varepsilon$ is a removal of node $u$ from the network. Any node $x$ not contained in the subtree of $u$ do not change its parent relation because the distance between $x$ and the root node $r$ is not modified (i.e., Predicate $\CParent(x)$ is not satisfied). Consider each edge $(u,v)$ between $u$ and its child $v$, we can apply the same argument described above for an edge removal. So only any node contained in the subtree connected by $u$ may change its parent.
\hspace*{\fill}$\Box$
\end{proof}

\subsection{Complexity analysis}
In the following we focus the complexity analysis of our algorithm in both static and dynamic networks.
Note that the original algorithm proposed in \cite{JT03} had no complexity analysis.
Interestingly, we prove that our extension has a zero time extra-cost with respect to the original solution. 
 
\begin{lemma}
\label{lem:time_complexity_static}
Starting from an arbitrary configuration, in at most $O(n^2)$ rounds a breadth first search tree is constructed by the algorithm in a static network.
\end{lemma}

\begin{proof}
To construct a spanning tree, the algorithm must remove all the cycles present in the starting configuration. So, we first analyze the number of rounds needed to remove a cycle.

To remove a cycle, a node of the cycle must change its parent to select a node out of the cycle, such a node is named a \emph{break node}. A node can change its parent using Rule $\RB$, but a break node executes Rule $\RB$ if the level of the new parent (out of the cycle) is lower than the level of the break node. Consider a break node $x$ and the neighbor $y$ of $x$ which must be selected as the new parent of $x$. We note $L_x$ and $L_y$ the level of $x$ and $y$ respectively. To select $y$ as its new parent and to break the cycle, $x$ must have its level $L_x$ such that $L_y<L_x$. In the cycle, a node corrects its level according to its parent by initiating a propagation of information with Rules $\RC$ and $\RD$. Thus the number of increments until we have $L_y<L_x$ is equal to $\lceil \frac{(L_x+1)-L_y}{|C|} \rceil$, with $|C|$ the size of the cycle $C$ to break. The propagation of information is in order of the size of $C$. Thus, $O((L_x+1)-L_y)$ rounds are needed to have $L_y<L_x$. Since we want to construct a breadth first search tree the level of a node cannot exceed $n$, with $n$ the size of the network. Thus, we consider that the level of a node is encoded using $\log n$ bits. The biggest value for $(L_x+1)-L_y$ is obtained when $L_y=1$ and therefore we have $(L_x+1)-L_y \leq n$.

Since the maximum number of possible cycles of a network is no more than $n/2$, obtained with cycles of size $2$, we have that in $O(n^2)$ all cycles are removed in the network and a spanning tree is constructed. In at most $O(D)$ additional rounds a breadth first tree is constructed, with $D$ the diameter of the network. Indeed, no cycle is created by the algorithm until reaching a legitimate configuration, since the algorithm guarantee the loop-free property.
\hspace*{\fill}$\Box$
\end{proof}

\begin{lemma}
\label{lem:time_complexity_dynamic}
Starting from an arbitrary configuration, in at most $O(n^2)$ rounds a breadth first search tree is constructed by the algorithm in a dynamic network.
\end{lemma}

\begin{proof}
In a dynamic network, for a node we can have the case where the edge leading to its parent or its parent is deleted from the network. When a node $x$ detects this case, $x$ executes Rule $\RF$ to find a new parent in the network. To accomplish this task, $x$ starts a propagation of information to increment its level since it has an incorrect level according to its parent (which is no more in the network).

We have two cases for the new parent selected by $x$. The first case is that the new parent of $x$ is a neighbor $y$ with level $L_y$ bigger than $x$'s level $L_x$. In this case, $x$ must increment its level to have the condition $L_y<L_x$. To obtain this condition, at most $L_y-L_x$ increments are needed, that is at most $n$ increments since we want to construct a breadth first tree and the level of a node is encoded using $\log n$ bits. The second case is that $x$ selects one of its children $u$ as its new parent, but to preserve the loop-free property $x$ can do this only when $u$ is no more a child of $x$. The worst case for $x$ is to wait that it has no more children if $u$ is its only child, that is the subtree of $x$ has disappeared. At most $n$ increments are needed to have that $x$ has an empty subtree.

In all cases, at most $n$ increments are needed and the number of rounds for a propagation of information is in the order of the size of the subtree of $x$, that is at most $n$. Thus, in at most $O(n^2)$ rounds $x$ finds a new parent in the network, then we can consider we are in the case of a static network and Lemma~\ref{lem:time_complexity_static} can be applied. Therefore, in at most $O(n^2)$ rounds a legitimate configuration is reached by the algorithm.
\hspace*{\fill}$\Box$
\end{proof}

\section{Super-stabilizing Loop-Free transformation scheme}


Our objective is to design a generic scheme for the construction of spanning trees considering any metric (not only metrics based on distances in the graph) 
with loop-free and super-stabilizing properties.
The idea is to extend an existing self-stabilizing spanning tree optimized for a given metric 
(e.g. MST, maximum degree spanning tree, max-flow tree etc) with super-stabilizing and loop-free 
properties via the composition with a spanning tree construction that already satisfies these properties.
Assume ${\mathcal M}$ be the predicate that captures the properties of the metric to be optimized.
Consider $\mathcal{A}$ the algorithm that outputs a self-stabilizing spanning tree and verifies ${\mathcal M}$. That is,
given a graph, $\mathcal{A}$ computes the set of edges $\mathcal{S_A}$ that satisfies ${\mathcal M}$ and is a spanning tree. 
Consider Algorithm $\mathcal{B}$ an algorithm that outputs a super-stabilizing and loop-free spanning tree $\mathcal{S_B}$. 
Ideally, if all edges in $\mathcal{S_A}$ are included in $\mathcal{S_B}$ then there is no need for further transformations. 
However, in most of the cases the two trees are not identical. Therefore, the idea of our methodology is very simple.
Algorithms $\mathcal A$ and $\mathcal B$ run such that the output of ${\mathcal A}$ defines the graph input for $\mathcal B$. That is,  
the neighborhood relation used by 
$\mathcal B$ is the initial graph filtered by ${\mathcal A}$ to satisfy the predicate $\mathcal M$.
The principal of this composition is already known in the literature as fair composition \cite{DIM93}. 
In our case the "slave" protocol is protocol ${\mathcal A}$ that outputs the set of edges input for the "master" protocol ${\mathcal B}$.
 
The following lemma direct consequence of the results proven in \cite{DIM93} 
guaranties the correctness of the composition.

\begin{lemma}
Let ${\mathcal M}$ be the predicate that captures the properties of the metric to be optimized.
Let $\mathcal{A}$ be an algorithm that outputs a self-stabilizing spanning tree that satisfies ${\mathcal M}$, $\mathcal{S_A}$.
Let $\mathcal{B}$ be a loop-free protocol that computes a spanning tree on the topology defined by 
$\mathcal{S_A}$ and super-stabilizing for a class of topology changes $\Lambda$.
The fair composition of ${\mathcal A}$ and ${\mathcal B}$ is a protocol that outputs a loop-free spanning tree 
that satisfies ${\mathcal M}$ and is super-stabilizing for $\Lambda$.
\end{lemma}
 
Note that our super-stabilizing loop-free BFS can be used as protocol ${\mathcal B}$ in the above composition.
The interesting property of the composition is that the time complexity will be maximum between $O(n^2)$ and the complexity 
time of the candidate to be transformed. Note that so far, the best time complexity of a spanning tree optimized for a given metric is $O(n^2)$ which 
leads to the conclusion that the composition does not alterate the time complexity of the candidate.

In the following, we specify the predicate $\mathcal{M}$ for two well known problems: max-flow trees and minimum degree spanning trees.

\paragraph{Case study 1: Maximum-flow tree}

The problem of constructing a maximum-flow tree from a given root node $r$ can be stated as follows. Given a weighted undirected graph $G=(V,E,w)$, the goal is to construct a spanning tree $T=(V,E_T)$ rooted at $r$, such that for every node $v \in V$ the path between $r$ and $v$ has the maximum flow. Formally, let $fw(v)=\min(fw(\parent_v),w(\parent_v,v))$ the flow for every node $v \in V$ in tree $\mathcal{T}$ and $mfw_v$ the maximum flow value of $v$ among all spanning trees of $G$ rooted at $r$. The maximum-flow tree problem is to compute a spanning tree $T$, such that $\forall v \in V, fw(v)=mfw_v$. The max flow tree problem has been studied \emph{e.g.} in \cite{GS99c}.
In this case, the graph $G_{\mathcal{S_A}}=(V_{\mathcal{S_A}},\mathcal{S_A})$ for the maximum-flow tree problem must satisfies the following predicate:
$$\mathcal{M} \equiv (|\mathcal{S_A}|=n-1) \wedge (V=V_{\mathcal{S_A}}) \wedge (\forall v \in V, fw(v)=\max\{\min(fw(u), w(u,v)) : u \in N(v)\}).$$

\paragraph{Case study 2: Minimum degree spanning tree}

Given an undirected graph $G=(V,E)$ with $|V|=n$, the minimum degree spanning tree problem is to construct a spanning tree $T=(V,E_T)$, 
such that the maximum degree of $T$ is minimum among all spanning trees of $G$. Formally, let $deg_{\mathcal{T}}(v)$ the degree of node 
$v \in V$ in the subgraph $\mathcal{T}$ and $deg(\mathcal{T})$ the maximum degree of subgraph $\mathcal{T}$ (i.e., $deg(\mathcal{T})=\max\{deg_{\mathcal{T}}(v): v \in V\}$). The minimum spanning tree problem is to compute a spanning tree $T$, such that $deg(T)=\min\{deg(T'): T' \mbox{ is a spanning tree of } G\}$.
A self-stabilizing solution for the minimum degree spanning tree algorithm has been proposed in \cite{BPR09}. If this solution plays the slave master in our transformation 
scheme then the graph $G_{\mathcal{S_A}}=(V_{\mathcal{S_A}},\mathcal{S_A})$  input for the BFS algorithm satisfy the following predicate:
$$\mathcal{M} \equiv (|\mathcal{S_A}|=n-1) \wedge (V=V_{\mathcal{S_A}}) \wedge deg(G_{\mathcal{S_A}})=\min\{deg(T'): T' \mbox{ a spanning tree of } G\}.$$

\section{Concluding remarks}

We presented a scheme for constructing loop-free and super-stabilizing protocol for universal tree metrics, without significant impact on the performance. There are several open questions raised by our work:

\begin{enumerate}
\item Decoupling various added properties (such as loop-freedom or super-stabilization) seems desirable. As a particular network setting may not need both properties and/or temporarily run in conditions where the network is essentially static, some complexity cost could be saved by removing uneeded properties. Of course, stripping our scheme can trivially result in a generic loop-free transformer or to a generic super-stabilizing transformer. Yet, modular design of features, as well as further enhancements (such as safe convergence~\cite{KM06c,KK08c}), seems an interesting path for future research.
\item The implementation of self-stabilizing protocols recently was helped by compilers that take as input guarded commands and provide as output actual source code for existing devices~\cite{DESAL}. Transformers such as this one would typically benefit programmers' toolboxes as they ease the reasoning by keeping the source code intricacies at a very high level. Actual implementation of our transformer into a programmer's toolbox is a challenging ingeneering task.
\end{enumerate}

\bibliography{biblio.bib}

\begin{thebibliography}{10}

\bibitem{Dij74}
Dijkstra, E.W.:
\newblock Self-stabilizing systems in spite of distributed control.
\newblock Communications of the ACM \textbf{17}(11) (1974)  643--644

\bibitem{Dolev00}
Dolev, S.:
\newblock Self-Stabilization.
\newblock MIT Press (2000)

\bibitem{T09bc}
Tixeuil, S.:
\newblock Self-stabilizing Algorithms.
\newblock Chapman \& Hall/CRC Applied Algorithms and Data Structures. In:
  Algorithms and Theory of Computation Handbook, Second Edition. CRC Press,
  Taylor \& Francis Group (November 2009)  26.1--26.45

\bibitem{GP93c}
Gopal, A.S., Perry, K.J.:
\newblock Unifying self-stabilization and fault-tolerance (preliminary
  version).
\newblock In: PODC. (1993)  195--206

\bibitem{AH93c}
Anagnostou, E., Hadzilacos, V.:
\newblock Tolerating transient and permanent failures (extended abstract).
\newblock In: WDAG. (1993)  174--188

\bibitem{DW97j}
Dolev, S., Welch, J.L.:
\newblock Wait-free clock synchronization.
\newblock Algorithmica \textbf{18}(4) (1997)  486--511

\bibitem{PT97j}
Papatriantafilou, M., Tsigas, P.:
\newblock On self-stabilizing wait-free clock synchronization.
\newblock Parallel Processing Letters \textbf{7}(3) (1997)  321--328

\bibitem{DW04j}
Dolev, S., Welch, J.L.:
\newblock Self-stabilizing clock synchronization in the presence of byzantine
  faults.
\newblock J. ACM \textbf{51}(5) (2004)  780--799

\bibitem{BDH08c}
Ben-Or, M., Dolev, D., Hoch, E.N.:
\newblock Fast self-stabilizing byzantine tolerant digital clock
  synchronization.
\newblock In: PODC. (2008)  385--394

\bibitem{MT06cb}
Masuzawa, T., Tixeuil, S.:
\newblock Bounding the impact of unbounded attacks in stabilization.
\newblock In: SSS. (2006)  440--453

\bibitem{MT07j}
Masuzawa, T., Tixeuil, S.:
\newblock Stabilizing link-coloration of arbitrary networks with unbounded
  byzantine faults.
\newblock International Journal of Principles and Applications of Information
  Science and Technology (PAIST) \textbf{1}(1) (December 2007)  1--13

\bibitem{DH97j}
Dolev, S., Herman, T.:
\newblock Superstabilizing protocols for dynamic distributed systems.
\newblock Chicago J. Theor. Comput. Sci. \textbf{1997} (1997)

\bibitem{H00j}
Herman, T.:
\newblock Superstabilizing mutual exclusion.
\newblock Distributed Computing \textbf{13}(1) (2000)  1--17

\bibitem{KUFM02j}
Katayama, Y., Ueda, E., Fujiwara, H., Masuzawa, T.:
\newblock A latency optimal superstabilizing mutual exclusion protocol in
  unidirectional rings.
\newblock J. Parallel Distrib. Comput. \textbf{62}(5) (2002)  865--884

\bibitem{CG02j}
Cobb, J.A., Gouda, M.G.:
\newblock Stabilization of general loop-free routing.
\newblock J. Parallel Distrib. Comput. \textbf{62}(5) (2002)  922--944

\bibitem{JT03}
Johnen, C., Tixeuil, S.:
\newblock Route preserving stabilization.
\newblock In: Self-Stabilizing Systems. (2003)  184--198

\bibitem{Aceves93}
Garcia-Luna-Aceves, J.J.:
\newblock Loop-free routing using diffusing computations.
\newblock IEEE/ACM Trans. Netw. \textbf{1}(1) (1993)  130--141

\bibitem{Gafni81}
Gafni, E.M., Bertsekas, P.:
\newblock Distributed algorithms for generating loop-free routes in networks
  with frequently changing topology.
\newblock IEEE Transactions on Communications \textbf{29} (1981)  11--18

\bibitem{Dolev_SuperStab}
Dolev, S., Herman, T.:
\newblock Superstabilizing protocols for dynamic distributed systems.
\newblock Chicago J. Theor. Comput. Sci. \textbf{1997} (1997)

\bibitem{BlinPRT09}
Blin, L., Potop-Butucaru, M., Rovedakis, S., Tixeuil, S.:
\newblock A new self-stabilizing minimum spanning tree construction with
  loop-free property.
\newblock In: DISC. (2009)  407--422

\bibitem{GS99c}
Gouda, M.G., Schneider, M.:
\newblock Stabilization of maximal metric trees.
\newblock In: WSS. (1999)  10--17

\bibitem{GuptaS03}
Gupta, S.K.S., Srimani, P.K.:
\newblock Self-stabilizing multicast protocols for ad hoc networks.
\newblock J. Parallel Distrib. Comput. \textbf{63}(1) (2003)  87--96

\bibitem{JT03-RR}
Johnen, C., Tixeuil, S.:
\newblock Route preserving stabilization.
\newblock Technical Report 1353, LRI, Universit\'e Paris-Sud XI (2003)

\bibitem{DIM93}
Dolev, S., Israeli, A., Moran, S.:
\newblock Self-stabilization of dynamic systems assuming only read/write
  atomicity.
\newblock Distributed Computing \textbf{7}(1) (1993)  3--16

\bibitem{BPR09}
Blin, L., Potop-Butucaru, M.G., Rovedakis, S.:
\newblock Self-stabilizing minimum-degree spanning tree within one from the
  optimal degree.
\newblock In: IPDPS. (2009)  1--11

\bibitem{KM06c}
Kakugawa, H., Masuzawa, T.:
\newblock A self-stabilizing minimal dominating set algorithm with safe
  convergence.
\newblock In: IPDPS. (2006)

\bibitem{KK08c}
Kamei, S., Kakugawa, H.:
\newblock A self-stabilizing approximation for the minimum connected dominating
  set with safe convergence.
\newblock In: OPODIS. (2008)  496--511

\bibitem{DESAL}
Dalton, A.R., McCartney, W.P., Dastidar, K.G., Hallstrom, J.O., Sridhar, N.,
  Herman, T., Leal, W., Arora, A., Gouda, M.G.:
\newblock Desal alpha: An implementation of the dynamic embedded
  sensor-actuator language.
\newblock In: ICCCN, IEEE (2008)  541--547

\end{thebibliography}
\bibliographystyle{splncs}

\end{document}